\documentclass{aa} % option [referee] to be inserted
\usepackage{graphicx}
\usepackage{natbib}
\usepackage[english]{babel}
\usepackage[T1]{fontenc}
\usepackage[latin1]{inputenc}
\usepackage[dvips]{epsfig}

\newcommand{\micron}{\mbox{\,${\mu}$m}}

\begin{document}

\title{IRAS 05358+3543: Multiple outflows at the earliest stages of massive star formation}
%\title{IRAS 05358+3543: Multiple and collimated outflows in massive star formation}

\author{H. Beuther\inst{1} \and P. Schilke\inst{1} \and F. Gueth\inst{1,2} \and M. McCaughrean\inst{3} \and M. Andersen\inst{3} \and T.K. Sridharan\inst{4} \and K.M. Menten\inst{1}}

\institute{Max-Planck-Institut f\"ur Radioastronomie, Auf dem H\"ugel 69, 53121 Bonn, Germany \and Institut de Radio Astronomie Millim\'etrique, 300 rue de la Piscine, 38406 Saint Martin d'H\`eres, France \and Astrophysikalisches Institut Potsdam, An der Sternwarte 16, 14482 Potsdam, Germany \and Harvard-Smithsonian Center for Astrophysics, 60 Garden Street, MS 78, Cambridge, MA 02138, USA}

\offprints{H. Beuther,\\ \email{beuther@mpifr-bonn.mpg.de}}

\date{Received ... /Accepted ...}

%%%%%%%%%%%%%%%%%%%%%%%%%%%%%%%%%%%%%%%%%%%%%%%%%%%%%%%%%%%%%%%%%%%%%%%%%
%              abstract
%%%%%%%%%%%%%%%%%%%%%%%%%%%%%%%%%%%%%%%%%%%%%%%%%%%%%%%%%%%%%%%%%%%%%%%%%

\abstract{We present a high-angular-resolution molecular line and millimeter 
continuum study of the massive star formation site {\it IRAS}
05358+3543. Observations with the Plateau de Bure Interferometer in CO
1--0, SiO 2--1 and H$^{13}$CO$^+$ 1--0 reveal at least three outflows
which cannot be separated in single-dish data. Observations at
millimeter and sub-millimeter wavelengths from the IRAM 30~m telescope
and the CSO provide additional information on the region. The most
remarkable feature is a highly collimated (collimation factor $\sim
10$) and massive ($>10$~M$_{\odot}$) bipolar outflow of $\sim1$~pc
length, which is part of a quadrupolar outflow system. The three observed
molecular outflows forming the {\it IRAS}~05358+3543 outflow system
resemble, in structure and collimation, those typical of low-mass
star-forming regions. They might therefore, just like low-mass
outflows, be explained by shock entrainment models of jets.  We
estimate a mass accretion rate of $\sim 10^{-4}$~M$_{\odot}$/yr,
sufficient to overcome the radiative pressure of the central object
and to build up a massive star, lending further support to the
hypothesis that massive star formation occurs similarly to low-mass
star formation, only with higher accretion rates and energetics. In
the millimeter continuum, we find three sources near the
center of the quadrupolar outflow, each with a mass of
75--100~M$_{\odot}$. These cores are associated with a complex region
of infrared reflection nebulosities and their embedded illuminating
sources. The molecular line data show
that SiO is found mostly in the outflows, whereas H$^{13}$CO$^+$
traces core-like structures, though likely with varying relative
abundances. Thermal CH$_3$OH comprises both features and can be
disentangled into a core-tracing component at the line center, and
wing emission following the outflows. A CO line-ratio study (using
data of the $J=1-0, 2-1~\&~6-5$ transitions) reveals local temperature
gradients.
}

\maketitle

%%%%%%%%%%%%%%%%%%%%%%%%%%%%%%%%%%%%%%%%%%%%%%%%%%%%%%%%%%%%%%%%%%%%%%%%%%%%
%                  Introduction
%%%%%%%%%%%%%%%%%%%%%%%%%%%%%%%%%%%%%%%%%%%%%%%%%%%%%%%%%%%%%%%%%%%%%%%%%%%%

\section{Introduction}

It is still not known whether the physical processes leading to
massive stars are similar to those of low-mass star formation, or
whether different processes are taking place. Classical star formation
scenarios predict moderate accretion rates around
$10^{-6}-10^{-5}$~M$_{\odot}$~yr$^{-1}$ \citep{shu 1977}, which are
incapable to overcome the radiation pressure for sources more massive
than approximately 10~M$_{\odot}$ \citep{wolfire 1987}. Observations
over the last decade showed that massive star formation occurs most
likely in a clustered mode, and theories were proposed that advocate
the coalescence of protostars in dense cluster centers to build up the
most massive sources \citep{bonnell 1998, stahler 2000}. In contrast,
according to other theories following the classical accretion
scenario, it is possible to form massive stars via enhanced accretion
and disks \citep{jijina 1996, norberg 2000,tan 2002,yorke 2002}.  For
more details on massive star formation see recent reviews, e.g.,
\citet{stahler 2000}, \citet{richer 2000}, \citet{churchwell 2000}, 
and \citet{kurtz 2000}.

Because molecular outflows give on large angular scales a wealth of
information about the innermost part of star-forming regions, in
recent years massive outflows have been investigated in more
detail. Such outflow studies seemed to indicate a lower degree of
collimation for massive outflows than known for low-mass
sources. Collimation factors between 1 and 2 were reported as compared
to low-mass outflows, where collimation factors up to 10 are found
(e.g., \citealt{richer 2000}). Based on those observations new outflow
scenarios were suggested, e.g., \citet{churchwell 2000} proposed that
massive outflows might be produced by deflection of infalling material
from the central protostar. But the analysis of the data has not taken
into account properly the spatial resolution of single-dish
observations.  Massive star formation sites are on the average far
more distant (a few kpc) than well known low-mass sources (a few
hundred pc), and thus their angular sizes are smaller in spite of
larger linear sizes. Recent statistical work on massive outflows with
$11''$ resolution by \citet{beuther 2001b} shows that the average
collimation --~even with this spatial resolution~-- is higher than
previously thought. Additionally, \citet{beuther 2001b} find that
low-mass correlations of outflow and core parameters continue up to
the high-mass regime, suggesting that similar star formation
processes are responsible for forming stars of all masses. But their
data also indicate that even higher angular resolution is needed to
disentangle real source and outflow structures, and thus detailed
studies with mm interferometers are needed.

There have been recent interferometric studies of massive outflow
sources, of which prominent examples are {\it IRAS}~20126+4104
\citep{cesaroni 1997,cesaroni 1999} or G192.16-3.82 
\citep{shepherd 1998,shepherd 1999}. While the outflow of 
{\it IRAS}~20126+4104 is rather collimated, G192.16-3.82 does not show
high degrees of collimation, and \citet{shepherd 1999} propose as
outflow mechanism a combination of a strong, wide-angle wind with a
weak jet.

In this paper, we present a detailed interferometric and single-dish
study at mm wavelengths of an extremely young and deeply embedded
high-mass protostellar object {\it IRAS}~05358+3543, which is also
known in the literature as S233IR. This source is part of a larger
sample presented and discussed by \citet{sridha} and \citet{beuther
2001a,beuther 2001b,beuther 2001c}.

{\it IRAS}~05358+3543 has different peak positions in the {\it IRAS}
12~$\mu$m and 100~$\mu$m bands. While the position in the Point Source
Catalog is based on the 12~$\mu$m emission due to an infrared cluster,
the 100~$\mu$m emission peaks approximately $40''$ away to the
north-east and traces the high-mass protostellar cluster we are
focusing on (Fig.\ \ref{mosaic}). As the mm dust continuum emission
coincides with the 100~$\mu$m peak, we will refer to that source as
{\it IRAS} 05358+3543mm. Based on {\it IRAS} infrared data, we
estimate the bolometric luminosity of the source to be $\sim
6300$~L$_{\odot}$, and the dust temperature to be around 47~K (Table
\ref{calc}, \citealt{sridha}). The first CO 1--0 observations of this 
region are reported by \citet{snell 1990}, who detected a 15
M$_{\odot}$ bipolar outflow and classify the source as a massive star
formation site. The estimated distance of 1.8~kpc is based on
observations of the nearby H{\sc ii} region S233 \citep{snell
1990}. As additional massive star formation and outflow signposts,
Class {\sc ii} CH$_3$OH \citep{menten 1991,sridha} and H$_2$O maser emission
\citep{tofani 1995,beuther 2001c} is observed towards 
{\it IRAS}~05358+3543mm. Additionally, OH maser emission is found in
the region with low spatial resolution \citep{wouterloot 1988}. As
reported by \citet{tofani 1995} and confirmed by \citet{sridha}, {\it
IRAS}~05358+3543mm does not show any 3.6~cm emission down to
1~mJy. Thus, based on the total luminosity of the region and a regular
initial mass function, it is likely that we are
witnessing a very early stage of massive star formation (see
\citealt{sridha}). The spectral type of the most massive object of the
cluster is likely between B2 and B1 ($10~{\rm{M_{\odot}}}\leq M \leq
13~{\rm{M_{\odot}}}$).

\citet{porras 2000}, \citet{yao 2000} and \citet{jiang 2001}
have published a series of near-infrared imaging and polarimetric
studies of this region, showing that there are two young embedded
stellar clusters, one associated with the 12~$\mu$m source (age
$\sim$\,3~Myr) and one with {\it IRAS}~05358+3543mm (age $\leq
2$~Myr). Narrow-band observations in the v=1--0 S(1) line of molecular
hydrogen at 2.122$\micron$ delineate at least two bipolar jets
emanating from the younger cluster, and polarimetric imaging
identifies the likely source of the outflows as being deeply embedded
and undetected at near-infrared wavelengths \citep{yao 2000,jiang
2001}. In an accompanying paper, \citet{mccaughrean 2001} present new,
more sensitive, and higher-spatial-resolution near- and mid-infrared
observations of the region, which reveal new information
concerning the shocked H$_2$ outflows and their driving sources.

%The outline of the paper is as follows: \S 2 describes the
%observations, and in \S 3 we present the
%observational results. In \S 4 we discuss and analyze the results,
%and a summary is presented in \S 5.

%%%%%%%%%%%%%%%%%%%%%%%%%%%%%%%%%%%%%%%%%%%%%%%%%%%%%%%%%%%%%%%%%%%%%%%%%%%%
%             Observation 
%%%%%%%%%%%%%%%%%%%%%%%%%%%%%%%%%%%%%%%%%%%%%%%%%%%%%%%%%%%%%%%%%%%%%%%%%%%%

\section{Observations}

We have observed {\it IRAS}~05358+3543mm in several tracers
with different telescopes.  Table \ref{parameter}
summarizes the observations described below.

\begin{table}[ht]
\caption{Observation parameters of spectral line observations; 
observatories are the 30~m at Pico Veleta (PV), the Plateau de Bure Interferometer (PdBI), the Caltech Submillimeter Observatory (CSO), and the Very Large Array (VLA); quoted system temperatures $T_{\rm{sys}}$ are average values, $\Delta v$ is the velocity resolution.}
\begin{tabular}{lrrrrr}
              & freq. & Obs & HPBW & $T_{\rm{sys}}$ & $\Delta v$ \\ &
              [GHz] & & [$''$] & [K] & [km/s] \\
\hline		              					
CO 1--0       & 115.27   & PV  & 22              & 120       & 0.2        \\
CO 1--0       & 115.27   & PdBI & $4.1\times 3.3$ & 300       & 0.4        \\
SiO 2--1      & 86.85    & PV  & 29              & 85        & 3.5        \\
SiO 2--1      & 86.85    & PdBI & $5.8\times 5.6$ & 150       & 0.3        \\
H$^{13}$CO$^+$& 86.75    & PV  & 29           & 85        & 3.5           \\
H$^{13}$CO$^+$& 86.75    & PdBI & $5.8\times 5.6$ & 150    & 0.3           \\
$^{13}$CO 1--0& 110.20   & PV  & 22              & 120       & 0.8       \\
CO 2--1       & 230.54   & PV  & 11              & 250       & 0.1        \\
CO 6--5       & 691.47   & CSO & 11              & 3000      & 0.02       \\
CH$_3$OH      & 241.79    & PV  & 11              & 250       & 0.1        \\
%H$_2$O        & 22.24    & VLA & 0.7             & 150       & 0.7        \\
%H$_2$         & $2.1\mu m$ & CA & 0.8      & --        & --         \\ 
\end{tabular}
\label{parameter}
\end{table}
 
\subsection{Plateau de Bure Interferometer (PdBI)}

{\it IRAS}~05358+3543mm was observed with the IRAM Plateau de Bure
millimeter array \citep{guilloteau 1992} between August and October
1999 in two frequency setups\footnote{The IRAM Plateau de Bure
Interferometer is supported by INSU/CNRS (France), MPG (Germany) and
IGN (Spain).}. Five 15~m antennas equipped with dual-frequency
receivers were used in three different configurations (4D1, 4C2 \&
4C2+N09) with baselines ranging from 24~m to 180~m. Only four antennas
were used during the August observations. Because of the poor summer
weather conditions, the quality of the 1~mm data was not satisfactory
in both setups, and they were only used for phase corrections. The
3~mm receivers were tuned to 115.27~GHz (USB) and 86.8~GHz (LSB) to
cover the CO 1--0 line, and the SiO 2-1 and H$^{13}$CO$^+$ 1--0 lines
simultanously. The phase noise was lower than 30$^{\circ}$ and
atmospheric phase correction based on the 1.3~mm total power was
applied. For continuum measurements, we placed two 160~MHz correlator
units in each band. Temporal fluctuations of amplitude and phase were
calibrated with frequent observations of the quasars 2145+067 and
0548+378. The amplitude scale was derived from measurements of MWC349
and CRL618. We estimate the final flux density accuracy to be $\sim
15\%$. The reference center is R.A.[J2000] 05:39:13.0 and Dec.[J2000]
35:45:54, and the $v_{\rm{LSR}}$ is $-17.6$~km~s$^{-1}$. Ten fields
(see Figure
\ref{mosaic}) were observed to cover the whole source, with the
exception of the first configuration which covered only the 6 eastern 
fields.

\begin{figure}
 \caption{The 1.2~mm bolometer image is plotted in grey-scale and
 dotted contours (the contour levels are 100~mJy~beam$^{-1}$ to
 1000~mJy~beam$^{-1}$ in steps of 100~mJy~beam$^{-1}$, in short
 100(100)1000~mJy~beam$^{-1}$), and the H$_2$ image
 \citep{mccaughrean 2001} overlayed in contours (arbitrary but linear
 units). The dashed circles present the observed PdBI mosaic (the four
 dashed-dotted western fields were observed in only two
 configurations). The circular H$_2$ feature at the right edge of the
 mosaic is due to the 12~$\mu$m {\it IRAS} source (black star). The
 white star shows the mm continuum peak and reference
 center of the mosaic (R.A.[J2000] 05:39:13.0 and Dec.[J2000]
 35:45:54). The offsets are given in arcsecs from the center of the
 mosaic.} \label{mosaic}
\end{figure}

\subsection{Pico Veleta (PV)}

{\it Line observations:}\\ To obtain large scale information about
{\it IRAS}~05358+3543mm, we used the IRAM 30~m telescope at Pico
Veleta in the Sierra Nevada, Spain in April 1999 and mapped the whole
region on-the-fly in the CO 2--1, CO 1--0, $^{13}$CO 1--0,
H$^{13}$CO$^+$ 1--0, SiO 2--1 and CH$_3$OH~$5_0-4_0(A+)$ lines (near
241.79~GHz). Due to the continuous movement of the telescope during
on-the-fly observations, spurious stripes in scanning direction (R.A.) 
are found in some maps (see Fig \ref{sd}). Except for CO 1--0
(observed later [May 2000] to add the short spacings, see \S
\ref{short}) all the other lines were observed simultaneously, which
guarantees the best alignment between the different maps. The
on-the-fly coverages were done twice with 2 sec integration time per
dump and a $4''$ grid (i.e., Nyquist-sampling at 241~GHz). As backends
we used the autocorrelator and two 1~MHz filterbanks.\\

\noindent {\it Bolometer observations:}\\ The 1.2~mm single-dish
continuum observations were conducted with the MAMBO array at the IRAM
30m telescope. For details on the observations and analysis
of this data see \citet{beuther 2001a}.

\subsection{Merging the interferometric and single-dish data}
\label{short}

The IRAM 30~m data (CO 1--0, SiO 2--1 and H$^{13}$CO$^+$ 1--0) 
were used to derive short-spacing information and thereby complement
the interferometric PdBI data. The algorithm used to derive the
visibilities corresponding to each pointing center of the mosaic is
described by \citet{gueth 1996}. The single-dish and interferometer
visibilities are subsequently processed together. Relative weighting
has been chosen to minimize the negative side-lobes in the resulting
dirty beam while keeping the highest angular resolution
possible. Again images were produced using natural weighting, then a
CLEAN-based deconvolution of the mosaic was performed. The final beam
sizes are $4.7''\times 3.8''$ (P.A. $51^{\circ}$) for CO 1--0, and
$5.9''\times 5.5''$ (P.A. $65^{\circ}$) for H$^{13}$CO$^+$ 1--0 and
SiO 2--1. The beam size for the 115~GHz continuum data, where only the
PdBI data were used, is $4''\times 3''$ (P.A. $46^{\circ}$).

\subsection{Caltech Submillimeter Observatory (CSO)}

We used the CSO\footnote{The CSO is operated by the California
Institute of Technology under funding from the National Science
Foundation, Grant No. AST 9980846.} 10.4 m telescope to obtain a CO
6--5 map of {\it IRAS}~05358+3543mm in the on-the-fly mode on
December, 15th 1999. At this frequency, the angular resolution of the
CSO is $11''$, well suited to be compared with the CO 2--1 data
of Pico Veleta. As backend we used the facility AOS. Further
observation parameters are given in Table
\ref{parameter}.

%%%%%%%%%%%%%%%%%%%%%%%%%%%%%%%%%%%%%%%%%%%%%%%%%%%%%%%%%%%%%%%%%%%%%%%%%%%%
%          Observational results
%%%%%%%%%%%%%%%%%%%%%%%%%%%%%%%%%%%%%%%%%%%%%%%%%%%%%%%%%%%%%%%%%%%%%%%%%%%%

\section{Observational results}

\subsection{Single-dish observations}

\subsubsection{Dust continuum emission}
\label{dust}
Figure 1 presents the 1.2~mm continuum image of {\it
IRAS}~05358+3543mm. The main peak is at the {\it IRAS}~100~$\mu$m peak
position and has more extended emission to the north-west and
the south-west. The latter sub-source is directly next to the
12~$\mu$m peak position, which is the center of a slightly older
cluster \citep{porras 2000} and can be interpreted as a remnant of
this older star-forming region. Additionally, there exists a
north-western elongation in the dust map, which will be shown to be
associated with molecular line emission (see section
\ref{single_dish_line}).

Assuming that the mm continuum is mainly produced by optically
thin dust emission with a grain emissivity index $\beta=2$,
\citet{beuther 2001a} estimate the total gas mass $M$ to $\sim
600~\rm{M}_{\odot}$ and the column density $N_{\rm{H_2}}$ to a few
times $10^{24}$~cm$^{-2}$ (Table
\ref{calc}), which shows that we are really dealing with a high-mass
star formation site. The column density converts to a visual
absorption magnitude $A_{\rm{v}}$ of $\sim 1000$
($A_{\rm{v}}=N_{\rm{H_2}}/0.94\times 10^{21}$, \citealt{frerking
1982}). If a central star has already ignited, its free-free emission
could be quenched by the infalling core \citep{walmsley 1995}.

\begin{table}[ht]
\caption{Physical parameters of the observed region from 1.2~mm 
continuum data ($M$ and $N_{\rm{H_2}}$, \S \ref{dust}), {\it IRAS}
($L$ and $T_{\rm{dust}}$, \citealt{sridha}) and CO 2--1
observations ($M_{\rm{out}}$, $\dot{M}_{\rm{out}}$ and dynamical age
$t$, \S \ref{single_dish_line})}
\begin{center}
\begin{tabular}{lr}
\hline
$M$ [M$_{\odot}$] & 610 \\ 
$N_{\rm{H_2}}$ [cm$^{-2}$] &
$4.0\times 10^{24}$ \\ $L$ [L$_{\odot}$] & 6300 \\
$T_{\rm{dust}}$ [K] & 47 \\
%$T_{\rm{hot}}$ [K] & 100  \\
$M_{\rm{out}}$ [M$_{\odot}$] & 20  \\
$\dot{M}_{\rm{out}}$ [M$_{\odot}$/yr] & $6\times 10^{-4}$ \\
$t$ [yr] & 36000\\
\hline
\end{tabular}
\end{center}
\label{calc}
\end{table}

\subsubsection{Molecular emission}
\label{single_dish_line}

\begin{figure*}
 \caption{IRAM 30~m single-dish observations. Each image shows the
 1.2~mm continuum observations in grey-scale and dotted contours
 (levels 100(100)1000~mJy~beam$^{-1}$) with the molecular line
 observations overlayed in contours (labeled in the top left
 corners). We show the integrated intensity of $^{13}$CO 1--0
 (v=[$-22,-9$]~km~s$^{-1}$, levels 32(4)52~K~km~s$^{-1}$),
 H$^{13}$CO$^+$ 1--0 (v=[$-26,-10$]~km~s$^{-1}$, levels
 0.4(0.6)5.2~K~km~s$^{-1}$) and SiO 2--1 (v=[$-26,-10$]~km~s$^{-1}$,
 levels 0.6(0.6)5.6~K~km~s$^{-1}$). The $^{12}$CO 2--1 map
 presents blue (solid contours, v=[$-32,-21$]~km~s$^{-1}$, levels
 13(9)40~K~km~s$^{-1}$) and red (dashed contours,
 v=[$-12,-4$]~km~s$^{-1}$, levels 13(9)40~K~km~s$^{-1}$) outflow
 emission. For thermal CH$_3$OH~$5_0-4_0(A+)$ emission at 241.79~GHz,
 we present one map for velocities near the line center
 (v=[$-17,-16$]~km~s$^{-1}$, levels 1.1(0.5)4.6~K~km~s$^{-1}$)
 and one map for the line wings (blue: solid lines
 v=[$-20,-23$]~km~s$^{-1}$, levels 2.1(0.5)4.6~K~km~s$^{-1}$;
 red: dashed lines v=[$-14,-12$] km~s$^{-1}$, levels
 1.1(0.5)1.6~K~km~s$^{-1}$).  The stretches in east-west direction,
 visible in the $^{12}$CO and $^{13}$CO maps, are artefacts of the
 on-the-fly observing mode. The beam size for each transition is shown
 in the bottom-right corners. \label{sd}}
\end{figure*}

Figure \ref{sd} presents single-dish data obtained at the 30~m
telescope on Pico Veleta. The $^{12}$CO~2--1 lines show self-absorption at
the line center (Fig. \ref{spectrum}), and blue and red wing emission
due to molecular outflows. Blue (v=[$-32,-21$] km~s$^{-1}$) and red
(v=[$-12,-4$] km~s$^{-1}$) $^{12}$CO wing emission maps were produced
by integrating the wing-part of the spectra. It is difficult to
disentangle bipolar structure in spite of the known outflow features
presented by \citet{porras 2000}, but the outflow lobes are centered
at the mm core. As outlined in \S \ref{pdb}, the inclination angle of
the outflow is low, and we use the mean intensities of the red and
blue line-wing maps in the spatial area known from the Plateau de Bure
observations for further outflow parameter determination. 
Opacity-corrected H$_2$ column densities $N_{\rm{b}}$ and $N_{\rm{r}}$
in both outflow lobes can be calculated by assuming a constant
$^{13}$CO/$^{12}$CO $2-1$ line wing ratio throughout the outflow
\citep{cabrit 1990}. \citet{choi 1993} found an average
$^{13}$CO/$^{12}$CO $2-1$ line wing ratio around 0.1 in 7 massive
star-forming regions, corresponding to a $\tau
(^{13}$CO~$2\to1)=0.1$. We adopt this value for our sample as well, and we
assume 30~K as average temperature in the outflow. The outflow mass
$M_{out}$, the dynamical timescale $t$, and the mass entrainment rate
$\dot{M}_{\rm{out}}$ are calculated via:
\begin{eqnarray*}
M_{\rm{out}} &=& (N_{\rm{b}} \times \rm{size_b} + N_{\rm{r}} \times
\rm{size_r})\ m_{\rm{H_2}} \\ t &=&
\frac{r}{(v_{\rm{max}_b}+v_{\rm{max}_r})/2}\\
\dot{M}_{\rm{out}} &=& \frac{M_{\rm{out}}}{t} \\
\end{eqnarray*} 
where $\rm{size_b}$ and $\rm{size_r}$ are the areas of the blue and
red outflow lobes, respectively, $m_{\rm{H_2}}$ the mass of the H$_2$
molecule, and $v_{\rm{max}_b}$ and $v_{\rm{max}_r}$ the maximum
velocities observed in each line wing. A more detailed description of
how the outflow parameters are determined is given in \citet{beuther
2001b}. According to \citet{cabrit 1990} derived masses are accurate
to a factor 2 to 4, whereas the error in the determination of the
dynamical parameters is higher, up to a factor 10. The derived total
outflow mass $M_{\rm{out}}$ is around 20~$M_{\odot}$ and the mass
entrainment rate of molecular gas $\dot{M}_{\rm{out}}$ approximately
$6\times 10^{-4}$~M$_{\odot}$yr$^{-1}$ (see Table
\ref{calc}). The latter value is derived dividing $M_{\rm{out}}$ by the
dynamical timescale $t$. As outlined in \citet{beuther 2001b}, such a
mass entrainment rate results in accretion rate estimates 
$\dot{M}_{\rm{accr}}$ around $10^{-4}$~M$_{\odot}$yr$^{-1}$. 
These are lower limits, because the outflow is not strongly inclined
to the plane of the sky (see section \ref{pdb}), which minimizes the
wing emission and by that the detectable (and separable) outflow
emission. In spite of the low accuracy, these values are high and
confirm the outflow to be massive.

\begin{figure}
 \caption{The $^{12}$CO 2--1 spectrum observed with the 30~m at the
 center position of the main mm core. The system velocity $v_{\rm{LSR}}$ is
 $-17.6$~km/s. \label{spectrum}}
\end{figure}

All other molecular line maps presented in Figure \ref{sd} peak at
positions that are clearly offset from the main mm core. This is in
contrast with the $^{12}$CO emission, where not only the wing emission
is centered on the main mm peak, but also the integrated emission (see
the $^{12}$CO~6--5 map in Figure \ref{single}). All data at Pico
Veleta were taken simultaneously, thus pointing errors cannot produce
these offsets. The $^{13}$CO~1--0, H$^{13}$CO$^+$, SiO and CH$_3$OH
images at the line center show a similar north-western elongation as
the mm continuum map. Low spatial as well as spectral resolution
of the H$^{13}$CO$^+$ and SiO data prevents further analysis of these
data, but the line wings of the CH$_3$OH data show a bipolar distribution
slightly shifted to the west with respect to the CO outflow,
indicating that the CH$_3$OH emission might trace a different
outflow in the west (see \S \ref{pdb}).

\begin{figure}
 \caption{The integrated CO 6--5 emission in contours
 (v=[--30,--3]~km~s$^{-1}$, levels 90(36)270~K~km~s$^{-1}$) is
 overlayed on the grey-scale 1.2~mm bolometer map (levels
 100(100)1000~mJy~beam$^{-1}$). The dashed lines show the western
 H$^{13}$CO$^+$ peak as outlined in \S \ref{h13co+} (1 channel at 
 $-13.5$~km/s is chosen, levels 25(25)225~mJy~beam$^{-1}$). The
 overlayed raster of arrows and markers outlines the different
 outflows and sub-sources as described in Figure \ref{obs}.}
\label{single}
\end{figure}

\subsubsection{Temperature gradients}
\label{temp}

\begin{figure}
\caption{Presented are in grey-scale the ratio maps of $^{12}\rm{CO}~6-5/2-1$ (left) and $^{12}\rm{CO}~1-0/2-1$ (right) separated in blue (lower) and red (upper) wing emission (as defined in \S\ref{single_dish_line}). In the blank white regions the signal to noise ratio is not sufficient to derive proper ratios. The contours show the 1.2~mm dust continuum map (levels 150(150)1050~mJy~beam$^{-1}$). The asterisk marks the hot-blue position (P1), the triangle the hot-red position (P2), and the circles the coldest position (P3). \label{co_ratios}}
\end{figure}

Multi-line studies of rotationally excited CO provide a good tool to
study temperature variations in molecular clouds. In thermalized gas,
where the local thermodynamical equilibrium (LTE) approximation
applies, the low-$J$ CO 1--0 and 2--1 lines trace cooler material
while the mid-$J$ CO 6--5 line is sensitive to warmer molecular gas
(e.g., \citealt{beuther 2000}, \citealt{hirano 2001}). To get an idea of the
temperature distribution in {\it IRAS}~05358+3543mm, we derived
$^{12}\rm{CO}~6-5/2-1$ and $^{12}\rm{CO}~1-0/2-1$ ratio maps for the
blue and red wing emission as defined in \S \ref{single_dish_line}
(Figure \ref{co_ratios}). All data are smoothed to the resolution of
the CO 1--0 transition ($22''$), even the 6--5/2--1 ratio map to
increase its signal-to-noise ratio.

In most regions rather uniform ratios around 0.5 are observed, but
each of the ratio maps has a prominent region with line ratios larger
than 1. The blue 6--5/2--1 ratios rise slightly east of the {\it IRAS}
Position (named P1 and marked by an asterisk in Figure
\ref{co_ratios}), and the red 6--5/2--1 ratios rise
south-east of the main mm peak (P2, triangle in Figure
\ref{co_ratios}), which is not prominent in the blue wings.
Contrasting to the increases in the 6--5/2--1 ratios around P1 and P2,
the 1--0/2--1 ratios are rather uniform around 0.5 throughout those
regions, but they rise --~independent of the line wings~-- around
$30''$ west of the mm peak, where the mm emission drops significantly
(P3, circle in Figure \ref{co_ratios}). The blue 6--5/2--1 ratios are
lowest there.

\begin{figure}
 \caption{CO ratios versus kinetic temperatures calculated via an LVG
 model for optical thick CO emission ($\tau (^{12}\rm{CO}\,
 2-1)\approx 6$). The left panel shows the $^{12}\rm{CO}~6-5/2-1$ ratio at
 densities of $10^5$~cm$^{-3}$, and the right panel presents the
 $^{12}\rm{CO}~1-0/2-1$ ratio at densities of
 $10^4$~cm$^{-3}$.\label{ttt}}
\end{figure}

As the average $^{13}$CO/$^{12}$CO $2-1=0.1$ line wing ratio (\S
\ref{single_dish_line}) corresponds to a $^{13}\rm{CO}\,2-1$ opacity
of $\approx 0.1$, the average $^{12}\rm{CO}\,2-1$ opacity in the wings
is about 6 \citep{langer 1990}. To quantify the temperatures in those
regions, we ran several Large Velocity Gradient models (LVG) for
optically thick CO emission ($\tau (^{12}\rm{CO}\,2-1) \approx
6$). The regions around P1 and P2 have average densities of a few
times $10^5$~cm$^{-3}$ \citep{beuther 2001a}, which is sufficient to
thermalize even the $J=6-5$ transition. To thermalize the CO~2--1
transition, densities of $10^4$~cm$^{-3}$ are sufficient, which is a
reasonable assumption at position P3. Fig. \ref{ttt} presents the
resulting line ratios versus the kinetic temperatures for the above
estimated densities, as calculated via the LVG
approximation. Obviously, increasing $^{12}\rm{CO}~1-0/2-1$ ratios do
indicate decreasing temperatures, whereas increasing
$^{12}\rm{CO}~6-5/2-1$ ratios are signposting enhanced temperatures in
such regions. The LVG calculations result in temperatures $\geq 80$~K
at P1 and P2, and temperatures below 20~K around P3.

\subsection{Interferometric observations}

\subsubsection{The main core at high resolution}

Zooming in the main single-dish mm core with the PdBI at 2.6~mm at a
spatial resolution of $4''\times 3''$ (7000$\times$5000~AU), it
resolves into three massive sub-sources with separations between $4''$
and $6''$. We label those three sources mm1, mm2 and mm3 (see Figure
\ref{out}). The major and minor core sizes were determined by fitting
two-dimensional Gaussian to the data, the integrated emission is
derived within approximately these areas. Assuming optically thin dust
emission, masses and column densities presented in Table \ref{mmm}
were calculated at 30~K (for more details on the calculations see
\citealt{beuther 2001a}).
%Source mm1
%is most likely the powering source of the main large scale outflow
%($\mathcal{A}$) and the high velocity outflow ($\mathcal{B}$) as
%outlined in the following section. 
Source mm1 coincides within less than $1''$ with the main mid-infrared
source (at $11.7~\mu$m) presented by \citet{mccaughrean 2001} as well
as with the deeply embedded source found by
\citet{yao 2000} and \citet{jiang 2001} in their K-band polarimetric
imaging study. 
%Source mm3 might be the center of the tentatively identified
%fourth outflow in east-west direction described at the end of the next section.

\begin{table}[h]
\caption{Parameters derived for the PdBI mm cores: positions, major and minor core sizes, peak and integrated intensities, and masses and column densities.\label{mmm}}
\begin{tabular}{lrrr}
                       & mm1 & mm2 & mm3 \\
\hline
R.A. [J2000.0] & 5:39:13.08   & 5:39:12.78 & 5:39:12.49 \\
Dec. [J2000.0] & 35:45:50.5    & 35:45:50.6  & 35:45:55.2\\
%position [$''$] & $0.8/-3.7$ & $-2.6/-3.5$ & $-6.7/0.8$ \\            
maj.$\times$min. [$''$] & $5.6 \times 4.5$ & $5.6 \times 4.1$ & $6.1 \times 4.1$\\
peak [mJy/beam]   & 23       & 16        & 16 \\
int. [mJy]        & 30       & 22        & 23 \\
$M$ [M$_{\odot}$]   & 100 & 73  & 77 \\
$N_{\rm{H_2}}$ [$10^{24}$cm$^{-2}$] & 5.1 & 3.5 & 3.5\\
\end{tabular}
\end{table}

Three H$_2$O maser features found by \citet{tofani 1995} are
associated with mm1, while one is associated with mm2. In 1999,
\citet{beuther 2001c} detected only one of the previously known H$_2$O
maser feature south of mm1 (Fig. \ref{out}).

The CH$_3$OH maser feature is located near the mm cores as well, but the
position is based on single-dish observations \citep{menten 1991} and
not accurate enough for a closer interpretation.

\citet{porras 2000} found 3 infrared sources (IR56, IR58 and IR93 in 
Figure \ref{out}) that are all offset from mm1 and mm2. \citet{yao
2000} and \citet{jiang 2001} then showed that the emission from IR58
and IR93 is highly polarized, and thus both are not independent
sources but reflection nebulae powered by mm1.

\begin{figure}
\caption{The grey-scale
 shows the 3~mm continuum emission (mm1, mm2, mm3; levels
 75(25)200~mJy~beam$^{-1}$) with the CO 1--0 high-velocity outflow
 ($\mathcal{B}$) overlayed in contours (blue: solid lines
 v=[--43,--29]~km~s$^{-1}$, levels
 0.5(1)3.5~Jy~beam$^{-1}$~km~s$^{-1}$; red: dashed lines
 v=[--9,0]~km~s$^{-1}$, levels 0.8(0.8)4~Jy~beam$^{-1}$~km~s$^{-1}$).}
\label{out}
\end{figure}

\subsubsection{Three (at least) bipolar outflows}
\label{pdb}

The most striking features of {\it IRAS}~05358+3543mm are the three
outflows observed with the Plateau de Bure Interferometer in CO and
SiO (Fig. \ref{obs}).\\

\noindent {\bf A large scale highly collimated CO outflow ($\mathcal{A}$)}\\ 
Figure \ref{obs}a shows a highly collimated large scale ($\approx
1$~pc) molecular outflow observed in CO in the east of the region.
We present the PdBI data without merging the single-dish data in
Figure \ref{obs}a, because the spatial filtering properties of the
interferometer help to isolate the flow from the ambient gas (the
combined data are presented in \S \ref{channel2}). To highlight the
differences between outflow ($\mathcal{A}$) and the high-velocity
outflow ($\mathcal{B}$), which is presented in the next paragraph, we
used slightly different velocity windows compared to the single-dish
image (Fig. \ref{sd}). Outflow ($\mathcal{A}$) is slightly bent and
oriented from north to south terminating in H$_2$ bow shocks. This
corresponds to one of the two outflows discussed by \citet{porras
2000} and by \citet{mccaughrean 2001} based on H$_2$ data. To quantify
the degree of collimation, we divide the length of the outflow by its
width, which results in a collimation factor of 10. A number of
emission peaks along the outflow show red and blue emission in CO and
SiO (Fig. \ref{obs}a,b), which is a typical feature for expanding bow
shocks near the plane of the sky. Thus, we conclude that the outflow
is not strongly inclined to the plane of the sky. \citet{porras 2000}
claim that IR93 is powering the outflow, but \citet{yao 2000} and
\citet{jiang 2001} show that IR93 is highly polarized and thus not a
separate source. Our mm data, the mid-infrared images by
\citet{mccaughrean 2001}, and the polarimetric images by \citet{yao
2000} and \citet{jiang 2001} strongly suggest that a protostar within
mm1 is the powering source of outflow ($\mathcal{A}$).\\

\noindent {\bf A high-velocity CO outflow ($\mathcal{B}$)}\\ A second bipolar 
outflow is seen in CO at high velocities relative to the system
velocity (Fig.\ \ref{out}, blue lobe v=[--43,--29]~km~s$^{-1}$ , red
lobe v=[--9,0]~km~s$^{-1}$). As Fig.\ \ref{out} shows, this outflow is
driven most likely by a source within the same protostellar
condensation as outflow ($\mathcal{A}$), and both outflows together
form a quadrupolar system, inclined by a position angle PA of $\approx
40^{\circ}$. \citet{mccaughrean 2001} differently speculate
that the second mid-infrared source $2''$ east might be the powering
source. Blue and red lobes are clearly resolved by the PdBI with the
red one to the south-east and the blue one to the north-west. The
high-velocity outflow corresponds to the second outflow detected in
H$_2$ emission by \citet{porras 2000} and further discussed by
\citet{mccaughrean 2001}. It has a
collimation factor around 6 (estimated from combined CO and H$_2$
data). At the tip of the south-eastern lobe, shocked SiO and H$_2$
emission is observed (Figure \ref{obs}a,b). This is also the
position of the warm region P2 (see section \ref{temp}), which
suggests that the temperature increase may be caused by
shock interaction of the outflow with the ambient medium.\\

\noindent {\bf A third outflow mainly observed in SiO ($\mathcal{C}$)}\\ 
Further to the west, a third outflow is observed mainly in SiO
2--1 on large scales in north-south direction (Fig.\ \ref{obs}b). Red
and blue lobes are resolved with three symmetric bullet-like features
in each lobe. But as the region is very complicated, we present two
alternate scenarios for the observed outflow features in this region.

{\bf (1) Outflow ($\mathcal{C}_{\rm{bent}}$):} The first hypothesis of
outflow structure is outlined by the two western arrows in
Fig. \ref{obs}. This way, the outflow shows the same bending as the
large scale CO outflow ($\mathcal{A}$) in the east, and it follows to
a large degree the mm emission as outlined in Fig.~\ref{single}. CO
emission is found there with the same bending structure (Figure
\ref{obs}a). The collimation factor of this outflow is around 3, and
the powering source is possibly the western H$^{13}$CO$^+$[3] core
(\S \ref{h13co+}).

{\bf (2) Outflow ($\mathcal{C}_{\rm{cavity}}$):} The other possible
interpretation of the observation is sketched by the two ellipses
presented in Fig. \ref{obs}. This way, the northern SiO emission is
not tracing the main part of the outflow, but rather the western
cavity wall. The northern elongated CO feature (Figure
\ref{obs}a), which is not accounted for in the previous scenario, 
could then represent the bow shock at the tip of the cavity. A problem
with this interpretation is that the SiO emission in the north is
mainly red-shifted with regard to the system velocity, whereas the
northern CO emission is on the blue side of the spectrum with a
difference of approximately 10~km~s$^{-1}$. A possible solution of
this discrepancy is that the outflow is in the plane of the sky at the
front side of the dense gas traced by the 1.2~mm dust continuum
emission. Thus, the SiO emission is produced at the backside of the
cavity, where the outflow interacts with the dense gas, and the
material moves away from the observer. The blue CO feature then could
be produced in the final expanding bow shock, when the outflow
interacts with less dense gas at the tip of the flow, which is in
front of the outflow and thus pushed towards the observer, producing
blue-shifted emission. In this interpretation, not H$^{13}$CO$^+$[3]
but rather the H$^{13}$CO$^+$ core [2] is the center of the outflow
(\S \ref{h13co+}), and the bending structure, especially of the
northern lobe, is less prominent. The collimation factor of this
structure is approximately 3 as well.

The southern part of the outflow seems to be less complicated and can
be interpreted in both scenarios. At the edge of the more evolved {\it
IRAS}~12~$\mu$m cluster, we find the hot blue region P1 described in
section \ref{temp}. The temperature increase of the outflowing gas in
this region could be caused by UV heating of the adjacent more evolved
cluster. We point out that the details of the morphological
interpretation of outflow ($\mathcal{C}$) have no bearing on the
overall energetics of the whole outflow system, which is dominated by
outflow ($\mathcal{A}$).\\

\noindent {\bf Outflow parameters}\\
We calculated the outflow parameters from the merged CO Plateau de
Bure and 30~m data with the same assumptions and within the same
velocity range as outlined in \S \ref{single_dish_line}. The derived
values (Table \ref{flowpdb}) agree reasonably well with the
single-dish results (Table \ref{calc}), and make us confident that the
orders of magnitude are correct. For this calculations we cannot
disentangle properly the contributions of the outflows ($\mathcal{A}$)
and ($\mathcal{B}$), thus the derived values include both
outflows. But from the spatial extent of both outflows, it becomes
clear that most of the outflowing mass and of the mass entrainment
rate is due to the large scale flow ($\mathcal{A}$). The value of
outflow ($\mathcal{C}$) is calculated for the morphologically bent
interpretation ($\mathcal{C}_{\rm{bent}}$).\\

\begin{table}[h]
\caption{Outflow parameters from the merged PdBI and Pico Veleta observations. 
\label{flowpdb}}
\begin{tabular}{lrrr}
           & $M_{\rm{out}}$ & $\dot{M}_{\rm{out}}$ & $t$ \\
           & [M$_{\odot}$]      & [$10^{-4}$M$_{\odot}$yr$^{-1}$] & [yr]\\
\hline
($\mathcal{A}$)+($\mathcal{B}$)    & 9.6                & 2.6  & 37000  \\
($\mathcal{C}_{\rm{bent}}$)        & 4.4                & 1.4              & 31000    \\
All emission & 16.9  & 4.6   & 37000 \\
\hline
\end{tabular}
\end{table}

\noindent {\sl Additional features}\\ Are there even more outflows? We 
tentatively identify one more outflow oriented in east-west
direction. Molecular line emission of CO and SiO shows extensions in
that direction, in the interferometric data as well as with the
single-dish observations (dotted east-west line in Figures
\ref{single}, \ref{obs}, and \ref{channel}). Furthermore, at the
western end of these molecular extensions, H$_2$ emission is
detected. One of the mm sources mm2 or mm3 might be the center of this
tentative flow. Finally, we note that even more outflows are indicated
in the H$_2$ data of \citet{mccaughrean 2001}.

\begin{figure*}
%\includegraphics[bb= 88.3723 74.7592 514.968 735.811, angle=-90,width=18cm]{pdb_col.ps}
%\hspace{-1.2cm} \psfig{angle=-90,width=21cm,file=/aux/pc048b/beuther/shortspacings/pdb.ps}
 \caption{Presented are the PdBI observations as contour overlays on
 the grey-scale H$_2$ data \citep{mccaughrean 2001}. {\bf (a)} 
 Outflow ($\mathcal{A}$): CO 1--0, red wing emission
 (v=[--14,--7]~km~s$^{-1}$, levels 5(10)95$\%$ from the peak
 intensity, PdB data only), and blue wing emission
 (v=[--30,--21]~km~s$^{-1}$, levels 10(10)90$\%$ from the peak
 intensity, PdB data only), the box marks the region shown in Figure
 \ref{out}. {\bf (b)} Outflow ($\mathcal{C}$): SiO 2--1, red wing
 emission ($\rm{v}=[-15.5,-8.5]$~km~s$^{-1}$, levels
 0.15(0.15)1.15~Jy~beam$^{-1}$~km~s$^{-1}$), and blue wing emission
 (v=[--28.5,--20.5]~km~s$^{-1}$, levels
 0.15(0.15)1.75~Jy~beam$^{-1}$~km~s$^{-1}$; PdBI data with short
 spacings from PV). {\bf (c)} H$^{13}$CO$^+$ 1--0, the integrated
 intensities are shown in green contours
 (v=[--20.5,--10.5]~km~s$^{-1}$, levels
 0.18(0.18)1.08~Jy~beam$^{-1}$~km~s$^{-1}$, PdBI data with short
 spacings from PV). The numbers in brackets label the three
 H$^{13}$CO$^+$ sources, the beams are shown each time at the
 bottom-right. In all images, the arrows outline the three main
 outflows, ($\mathcal{A}$): long north-south arrows in the east,
 ($\mathcal{B}$): short arrows in the east, and
 ($\mathcal{C}_{\rm{bent}}$): long arrows in the west. The ellipses
 outline the second interpretation of the western outflow
 ($\mathcal{C}_{\rm{cavity}}$), and the dotted east-west line shows
 the speculative forth outflow. The three squares represent the mm
 sources, the diamonds locate the H$^{13}$CO$^+$ peaks [2] and [3],
 and the triangle marks the 12~$\mu$m {\it IRAS}~position.}
 \label{obs}
\end{figure*}

\subsubsection{H$^{13}$CO$^+$ emission}
\label{h13co+}

The PdBI data of H$^{13}$CO$^+$ 1--0 show three peaks in
east-west orientation, and more extended emission to the north and
south. One H$^{13}$CO$^+$ peak (H$^{13}$CO$^+$[1]) is associated with
the three mm continuum peaks (mm1 to mm3). The other
H$^{13}$CO$^+$ peaks [2] and [3] are both close to the center of
outflow ($\mathcal{C}$), but we cannot determine with certainty, which
of the two is the powering source of this flow (\S \ref{pdb}). 
Assuming local thermodynamic equilibrium, an HCO$^+$ abundance of
$1\times 10^{-9}$
\citep{vdishoek 1993} and a C to $^{13}$C ratio of 67
\citep{langer 1990}, we can estimate the approximate masses of the
H$^{13}$CO$^+$ clumps, which are listed in Table \ref{h13}. The
temperatures decrease to the west according to the line ratio results
found in \S \ref{temp}.

\begin{table}[h]
\caption{Masses of the H$^{13}$CO$^+$ clumps at the given temperatures. 
The temperatures decrease to the west (\S \ref{temp}). \label{h13}}
\begin{tabular}{lccc}
H$^{13}$CO$^+$ & [1] & [2] & [3] \\ 
\hline
mass [M$_{\odot}$] & 19 & 6 & 8\\
T    [K] & 30 & 20 & 15
\end{tabular}
\end{table}

The three H$^{13}$CO$^+$ clumps are aligned in east-west
direction with similar fluxes in each clump. Morphologically, this is
different from the mm continuum emission which decreases 
towards the west (see Fig. \ref{mosaic}). Therefore, the
H$^{13}$CO$^+$ peaks correspond only weakly to column density and mass
concentrations traced by the mm continuum. As we do not have other
high-resolution molecular line observations, it is difficult to
discriminate between abundance and density variations, and it is
probable that the H$^{13}$CO$^+$ peaks [2] and [3] are caused by an
interplay of different processes. The mass of H$^{13}$CO$^+$[1] is
about an order of magnitude below the value we derive from the mm
continuum (Tables \ref{calc} \& \ref{mmm}). Thus, in this region
H$^{13}$CO$^+$ is likely depleted. Remarkably, we detect no mm source
near H$^{13}$CO$^+$[2] and [3] and therefore at the center of outflow
($\mathcal{C}$). The 2.6~mm PdBI data show at the H$^{13}$CO$^+$[3]
position just a peak at the 2$\sigma$ level, and our 3$\sigma$ mm
sensitivity corresponds to a mass limit of $\sim 50$~M$_{\odot}$ (at
15~K). Even by a factor 5 enhanced H$^{13}$CO$^+$ abundances result in
lower mass estimates based on the H$^{13}$CO$^+$ data, and the clump
stays undetected in the mm continuum with our sensitivity limit. This
is also the region P3, where the temperature decreases significantly
according to the CO $1-0$/$2-1$ ratios (\S \ref{temp}), and where only
weak emission is observed in the integrated CO $6-5$ map
(Fig.~\ref{single}).

\subsubsection{Channel maps of the CO data}
\label{channel2}

In Figure \ref{obs}a we show for clarity the CO~1--0 data from
the interferometer only, since the large-scale flow is more pronounced than
in the merged data. Figure
\ref{channel} now presents a channel map of the Plateau de Bure data
merged with short spacings obtained at Pico Veleta. The eastern large
scale outflow is still visible (e.g., channels at $-25.5$, $-23.5$,
$-7.5$ and $-5.5$ [km s$^{-1}$]), and new features appear
by adding the short spacings due to bright extended emission at
velocities close to the system velocity.

The ring-like structures seen in a number of channels are no
artefacts caused by the merging with the interferometric data, because
the ring is also seen in the $^{12}$CO~6--5 single-dish data (Fig.
\ref{single}).  At $-21.5$~km~$^{-1}$, the ring-like structure seems to
surround the position of the H$^{13}$CO$^+$[3] peak, which suggests
that self absorption could cause the ring-like structure. But as the
center of this structure moves spatially between the different
channels, the ring might be produced by 
different overlapping outflows. The east-west structure at $-23.5$~km
s$^{-1}$ has the same orientation as the putative fourth flow,
which supports our idea of another outflow.

\begin{figure*}
%\includegraphics[bb=64.2436 40.5915 464.915 709.354 , angle=-90,width=18cm]{05358_channel.ps} 
%\psfig{angle=-90,width=18cm,file=/aux/pc048b/beuther/shortspacings/05358_channel.ps}
%\vspace{-2cm}
 \caption{Channel map of the region in CO 1--0, Plateau de Bure data
 merged with short spacings from Pico Veleta. The grey-scale levels
 are between 0 and 10~Jy~beam$^{-1}$~km~s$^{-1}$, and the contours
 range between 0.1 and 5.1 (by 1) Jy~beam$^{-1}$~km~s$^{-1}$. The
 bottom-right corner shows the centroid velocity of each channel. The
 overlayed arrows and markers in two channels are the same as in
 Fig. \ref{pdb}.}  \label{channel}
\end{figure*}

\subsection{Interpretation of single-dish maps with PdBI data}

Some of the puzzling single-dish features outlined in section
\ref{single_dish_line} can be explained with the high-resolution data.

Being near the plane of the sky, outflow ($\mathcal{A}$) is difficult
to resolve from the ambient gas in the single-dish data, and the
outflows are far better visible with the Plateau de Bure
Interferometer alone. Its higher spatial resolution is capable of resolving
different flows spatially, whereas the interferometric feature of
filtering out large scale uniform emission makes outflow
($\mathcal{A}$) a very prominent structure (Fig.
\ref{obs}a).

Interestingly, the offsets between the molecular line peaks of
H$^{13}$CO$^+$, SiO, and CH$_3$OH, and the 1.2~mm continuum peak are
based on different physical processes in spite of their similar
spatial distribution in the single-dish observation (Fig.\
\ref{sd}). In the case of H$^{13}$CO$^+$, the PdBI data resolve the
single-dish map into three sub-sources with an extension along the
north western 1.2~mm continuum ridge, which makes H$^{13}$CO$^+$ being
at least partly a tracer of core-structure, though likely with varying
relative abundances (\S \ref{h13co+}). In contrast, SiO is only found
in the outflows, but because {\it IRAS}~05358+3543mm exhibits at least
three outflows, the single-dish map looks very similar to the
H$^{13}$CO$^+$ image.

The situation is different with regard to the thermal
CH$_3$OH emission at 241~GHz. While the line center emission
corresponds to the dust emission, the wing emission seems to be
associated with outflow ($\mathcal{C}$). From these data it appears
that CH$_3$OH is a tracer of outflows as well as of core emission.

%%%%%%%%%%%%%%%%%%%%%%%%%%%%%%%%%%%%%%%%%%%%%%%%%%%%%%%%%%%%%%%%%%%%%%%%%%
%                            Discussion
%%%%%%%%%%%%%%%%%%%%%%%%%%%%%%%%%%%%%%%%%%%%%%%%%%%%%%%%%%%%%%%%%%%%%%%%%%

\section{Discussion}  
\label{discussion}

\subsection{The 1~pc, highly collimated outflow ($\mathcal{A}$)}
\label{flow}

The eastern large scale outflow ($\mathcal{A}$) in {\it
IRAS}~05358+3543mm is the first example of a highly collimated,
jet-like, bipolar and massive outflow with an extension of $\approx$
1~pc. The collimation factor 10 is as high as the highest found in
low-mass flows. High degrees of collimation are difficult to explain
by stellar winds or deflection of infalling matter.  Outflow models
involving highly collimated jets entraining the surrounding material
are much more likely \citep{cabrit 1997}. \citet{churchwell 2000}
argues that the predictions of shock-entrainment models for the amount
of material that can be entrained are far too low to account for
several tens of solar masses found in massive outflows, as it is the
case for {\it IRAS}~05358+3543mm (Table \ref{calc}). But he also
stresses the caveats in these estimates, because they are highly
sensible to different assumptions, especially the entrainment
efficiency in the dense interstellar medium as found in massive star
formation sites. It is well possible that the entrainment efficiency
rises significantly in the very dense interstellar medium, and
higher efficiencies could easily account for the observed high outflow
masses found in massive star-forming regions. Therefore, outflow
($\mathcal{A}$) links massive outflow phenomena to scenarios already
known from low-mass star formation. It strongly suggests that
entrainment by bow shocks propagating in a collimated jet also has
an important role in the formation of massive outflows.

\subsection{The high-velocity outflow ($\mathcal{B}$) and the quadrupolar 
structure}

Emanating at a PA of $\approx 40^{\circ}$, most likely from the same
mm core mm1 (Fig.\ \ref{out}), the less extended high-velocity outflow
($\mathcal{B}$) is clearly distinct from the large scale outflow
($\mathcal{A}$). Different quadrupolar outflow mechanisms are
discussed in the literature (for a brief compilation and discussion
see \citealt{gueth 2001}), and the most appealing explanation in the
case of {\it IRAS}~05358+3543mm seems to be that the two outflows are
produced independently by adjacent protostars inside the same mm
condensation mm1. Being parts of one and the same outflow is unlikely,
mainly because the large scale outflow does not show any elongation at
a PA of $40^{\circ}$, which would be expected if the quadrupolar
structure was due to deflected or precessing material. Assuming half
of the HPBW ($\approx 1.75''\approx 3100$~AU) at a distance of 1.8~kpc
as the the maximal separation of the two powering sources, this is
well within the range of typical binary separation.

\subsection{The SiO outflow ($\mathcal{C}$)}

In the bent interpretation, the SiO outflow
($\mathcal{C}_{\rm{bent}}$) closely follows the north-south elongation
of the dust continuum emission. It is slightly less collimated than
the large CO outflow ($\mathcal{A}$), and blue and red lobes are
separated in SiO as well as in CO. The northern and southern lobes
show three peaks each that are symmetric with respect to the assumed
origin, suggesting three different ejection events. In contrast to
this, the cavity interpretation of outflow
($\mathcal{C}_{\rm{cavity}}$) does not imply that the outflow has to
follow the core structure, but rather that SiO is only observed in
that part of the flow where the densities are high enough to excite
SiO sufficiently. As already mentioned, outflow
($\mathcal{C}_{\rm{cavity}}$) then is likely to be located at the
front side of the core outlined in Fig. \ref{mosaic}. The bullet-like
structures and their interpretation as three ejection events are
independent of the exact scenarios.

It is a rather intriguing fact that SiO, which forms in shocks from
dust disruption \citep{schilke 1997}, is only found in regions with
strong dust emission (Figure \ref{single}). Outflow ($\mathcal{C}$)
follows to a large degree the north-south dust filament, while the
eastern large scale CO outflow leaves the dense core very soon. Thus,
throughout most parts of outflow ($\mathcal{A}$), column densities are
low and the gas densities there are not high enough to excite the SiO
sufficiently. Another reason may be that that C-shocks (shock
velocities $v_s\leq 50$~km s$^{-1}$) are needed to produce SiO
\citep{schilke 1997}. In the less massive outflow ($\mathcal{C}$)
C-shocks are likely to be more common than in the more massive outflow
($\mathcal{A}$), where higher jet velocities are expected
($>500$~km~s$^{-1}$, e.g., \citealt{eisloeffel 2000}) and shocks are
more likely of J-type ($v_s\geq 50$~km s$^{-1}$), dissociating the
molecular material. These different scenarios have to be checked for a
larger sample.

\subsection{The large scale bending}

Assuming the bent interpretation of outflow
($\mathcal{C}_{\rm{bent}}$), both large scale outflows and the mm dust
core show the same bent morphology. For the cavity interpretation of
outflow ($\mathcal{C}_{\rm{cavity}}$), still outflow ($\mathcal{A}$),
the southern lobe of outflow ($\mathcal{C}_{\rm{cavity}}$), and the mm
dust emission exhibit the same bent structure. Possible bending
mechanisms for protostellar jets are discussed, e.g., in Fendt
\& Zinnecker (1998).

Internal bending scenarios on small scales seem to be ruled out in the
case of {\it IRAS}~05358+3543mm, because it is not likely that in both
large scale outflows exactly the same small scale physics takes place
(e.g., acceleration of the jet source by a binary component, or
precession of an accretion disk in a binary system), and additionally
the dust core follows a similar morphology.

Considering the whole source structure, external effects seem to be
more plausible to explain the structure of this massive star-forming
site. On larger scales (4 to 10~pc) other H{\sc ii} regions are found:
in the north-east S233 (at a distance of $\approx 4$~pc), in the west
S235 (at a distance of $\approx 10$~pc), in the north-west S231
(distance $\approx 4$~pc) and a bit further away S232 (distance
$\approx 15$~pc, for large scale images see, e.g.,
\citealt{porras 2000}). As a whole, the region contains different
massive star formation sites at the edge of {\it IRAS}~05358+3543mm,
and the most likely explanation of the overall bending of this
youngest site is due to energy input (UV radiation as well as stellar
winds) from the more evolved star forming regions in the
vicinity. Thus, we are possibly observing an example of sequential
star formation and interactions of regions of different ages.

\subsection{Outflows from young stellar objects of all masses}

A key question triggering this observational study was whether massive
outflows are generated by different physical mechanism than their
low-mass counterparts. We therefore compare IRAS~05358+3543mm with
other outflow sources of different core masses.

{\it A low-mass object:} a prototype of a low-mass outflow is HH211
emanating from a dust condensation of $\sim 0.2$~M$_{\odot}$
\citep{gueth 1999}. It includes a highly collimated, high-velocity
molecular jet surrounded by a less collimated cavity-like outflow at
lower velocity. A shock-entrainment model nicely explains the outflow
structure.

{\it An intermediate-mass object:} \citet{gueth 2001} recently
observed the intermediate-mass young stellar object powering HH288
(core mass between 6 and 30~M$_{\odot}$). This outflow is quadrupolar
and most likely due to two powering sources not separable by the
resolution of the observation. But in spite of being quadrupolar, the
whole system can be explained by two outflows with shock-entrained
material similar to HH211.

{\it A high-mass object:} our observations of {\it IRAS}~05358+3543mm
show an example of a massive star-forming cluster. High-resolution
Plateau de Bure observations resolve the single-dish features into a
number of different outflows. So far, just a few massive outflows have
been mapped with high spatial resolution (see Introduction), and we
stress that the large scale eastern outflow is the first massive
outflow observed with such a high degree of collimation (collimation
factor $\sim 10$). We estimate the accretion rate to be
$10^{-4}$~M$_{\odot}$yr$^{-1}$, and \citet{beuther 2001b} recently
showed that mass entrainment rates of massive star-forming regions are
usually of that order. This is high enough to overcome the radiation
pressure of the central object and build up more massive stars
\citep{wolfire 1987,jijina 1996,norberg 2000,tan 2002,yorke 2002}. 
Additionally, jet models require disks, which are believed to be
observed in a few massive objects (e.g., {\it IRAS}~20126+4104,
\citealt{cesaroni 1997},
\citealt{zhang 1998}; {\it IRAS}~23385+6053 \citealt{molinari
1998}). Thus, to explain the outflow features observed in {\it
IRAS}~05358+3543mm, shock-entrainment models are sufficient (e.g.,
\citealt{gueth 1999}, \citealt{richer 2000}, and \citealt{cabrit 1997}),
and no other formation mechanism is needed. It cannot be ruled out
that in other sources different physical mechanisms are taking place,
but these observations indicate that other massive star formation
sites of rather chaotic appearance might be disentangled into simpler
structures if they are observed at higher angular resolution.

%%%%%%%%%%%%%%%%%%%%%%%%%%%%%%%%%%%%%%%%%%%%%%%%%%%%%%%%%%%%%%%%%%%%%%%%%
% SUMMARY
%%%%%%%%%%%%%%%%%%%%%%%%%%%%%%%%%%%%%%%%%%%%%%%%%%%%%%%%%%%%%%%%%%%%%%%%%

\section{Summary}

{\it IRAS}~05358+3543mm is the first example of a massive
($>10$~M$_{\odot}$) bipolar outflow with a high degree of collimation
on scales of 1~pc (collimation factor $\sim 10$). High-angular-resolution
observations with the Plateau de Bure Interferometer resolve the
single-dish observations into at least three different outflows. Two
of them form a quadrupolar system, most likely emanating from adjacent
protostars within the same mm core. Our data resolve three massive
mm cores (between 100~M$_{\odot}$ and 75~M$_{\odot}$) at the center of
the quadrupolar outflow. 

The data suggest that the physical processes associated with this
massive outflow are similar to those driving their low-mass
counterparts. It is likely that many massive star formation sites can
be shown by interferometric high-resolution observations to be
composed of basic features known from low-mass star formation. The
accretion rate is high ($\sim 10^{-4}$~M$_{\odot}$yr$^{-1}$) and
consistent with disk-accretion scenarios explaining the formation of
massive stars (e.g., \citealt{wolfire 1987}, \citealt{jijina 1996},
\citealt{norberg 2000}, \citealt{tan 2002}, and \citealt{yorke 2002}).

The overall distribution of H$^{13}$CO$^+$, SiO and thermal CH$_3$OH
looks very similar, but the higher resolution PdBI observations show
varying morphologies traced by the different lines. While SiO is
observed mainly in the outflows, H$^{13}$CO$^+$ traces core-like
structures, which do not coincide exactly with the dust cores because
of varying relative H$^{13}$CO$^+$ abundances. Finally, CH$_3$OH can
be decomposed into a core tracing component at the line center and
wing emission tracing the outflows.

Ratio maps between CO 6--5, 2--1 and 1--0 reveal local temperature
gradients. At the tip of the high-velocity outflow ($\mathcal{B}$), we
find a temperature increase ($\geq 80$~K) caused by shock-interaction
of the outflow with the ambient medium. Additionally, the southern
lobe of the third outflow ($\mathcal{C}$) is much warmer (again $\geq
80$~K) than the rest of the outflow, possibly due to UV heating of a
close by and more evolved cluster. Contrasting to these
temperature increases, the possible center H$^{13}$CO$^+$[3] of
outflow ($\mathcal{C}$) is cold, below 15~K.

%%%%%%%%%%%%%%%%%%%%%%%%%%%%%%%%%%%%%%%%%%%%%%%%%%%%%%%%%%%%%%%%%%%%%%%%%%
% acknoledgements
%%%%%%%%%%%%%%%%%%%%%%%%%%%%%%%%%%%%%%%%%%%%%%%%%%%%%%%%%%%%%%%%%%%%%%%%%%

\begin{acknowledgements} 
We like to thank an unknown referee for helpful comments on the initial
draft of this paper. H. Beuther is supported by the {\it Deutsche
Forschungsgemeinschaft, DFG} project number SPP 471.
\end{acknowledgements}

%%%%%%%%%%%%%%%%%%%%%%%%%%%%%%%%%%%%%%%%%%%%%%%%%%%%%%%%%%%%%%%%%%%%%%%%%%
% bibliography
%%%%%%%%%%%%%%%%%%%%%%%%%%%%%%%%%%%%%%%%%%%%%%%%%%%%%%%%%%%%%%%%%%%%%%%%%%

\end{document}